\documentclass[preprint,12pt,3p]{elsarticle}

\usepackage[ruled]{algorithm2e}
\usepackage{algpseudocode}
\usepackage{amsthm} 
\usepackage{amsmath}
\usepackage{amssymb}
\usepackage{rotating}
\usepackage{siunitx}
\usepackage{subfigure}
\usepackage{float}
\usepackage{verbatim}

\newtheorem{theorem}{Theorem}[section]

\newtheorem{lemma}[theorem]{Lemma}
\newtheorem{corollary}[theorem]{Corollary}
\newtheorem{definition}[theorem]{Definition}

\newtheorem{remark}[theorem]{Remark}

\begin{document}


\begin{frontmatter}

\title{ On the GI-Completeness of a Sorting Networks Isomorphism }

\author{Martin Marinov\fnref{label:Martin}}
\ead{marinovm@tcd.ie}
\fntext[label:Martin]{Corresponding author. Dept. of Computer Science, Trinity College Dublin, Ireland. Work supported by the Irish Research Council (IRC).}

\author{David Gregg\fnref{label:David}}
\ead{dgregg@cs.tcd.ie}
\fntext[label:David]{Lero, Trinity College Dublin.}

\begin{abstract}
Our main research interest is optimizing $n$-input sorting networks --- a mathematical object oblivious to the order of the input data which always performs the same set of pre-determined operations to produce a sorted list of $n$ numbers. In the early $2000$'s a sorting/comparator network isomorphism and normalization is presented by Choi and Moon. This is used to substantially reduce the search space for optimal $n$-input sorting networks by considering only representative networks up to the isomorphism. Choi and Moon prove the computational complexity of checking whether two $n$-input networks are isomorphic to be polynomially reducible to the bounded valence graph isomorphism (GI) problem.  In 2013, Bundala and Zavodny described a new sorting network relation (subitemset isomorphism) which is `superior' to that of Choi and Moon --- any networks that are CM (Choi and Moon) isomorphic are also BZ (Bundala and Zavodny) subitemset isomorphic but the converse is not true in the general case. Bundala and Zavodny's sorting network subitemset isomorphism drastically reduces the search space for optimal sorting networks in comparison to previous methods. Their (BZ) isomorphism is at the core of their computer-assisted proof for depth optimality of $n$-input sorting networks for $11 \leq n \leq 16$ and also Codish~et~al's comparator optimality computer-assisted proof for nine and ten-input sorting networks.

The subitemset isomorphism problem is really important and there are excellent practical solutions described in the literature. However, the computational complexity analysis and classification of the BZ subitemset isomorphism problem is currently an open problem. In this paper we prove that checking whether two sorting networks are BZ isomorphic to each other is GI-Complete; the general GI (Graph Isomorphism) problem is known to be in NP and LWPP, but widely believed to be neither P nor NP-Complete; recent research suggests that the problem is in QP. Moreover, we state the BZ sorting network isomorphism problem as a general isomorphism problem on itemsets --- because every sorting network is represented by Bundala and Zavodny as an itemset. The complexity classification presented in this paper applies sorting networks, as well as the general itemset isomorphism problem. The main consequence of our work is that currently no polynomial-time algorithm exists for solving the BZ sorting network subitemset isomorphism problem; however the CM sorting network isomorphism problem can be efficiently solved in polynomial time.

\end{abstract}

\begin{keyword}
Itemset Isomorphism \sep Graph Isomorphism \sep GI-Complete \sep GI-Hard \sep Optimal Sorting Networks
\end{keyword}

\end{frontmatter}

\section{Introduction}
\label{sec:intro}

\subsection{Structure}
\label{sec:intro:struct}

We have structured the presentation of our work in the following manner. First, we give a brief introduction to the sorting network optimization problem to describe one real world instance of the (generic) problem tackled in this paper. Next, we give a formal description of the Subitemset Isomorphism (SI) problem together with necessary terminology. Then, we give a summary of the related work on the complexity classification of the problem. In Section~\ref{sec:complexity}, we present our main result by proving the Itemset Isomorphism (II) problem is GI-Complete; an immediate consequence is that the SI problem is GI-Hard. In \ref{sec:examples:GI<II} and \ref{sec:examples:II<HGI} we present a set of examples that illustrate main points of the rather technical complexity classification proofs. We conclude by presenting a brief summary of our work and discuss possibilities for future contributions.

\subsection{Terminology}
\label{sec:intro:term}

\begin{figure} [t]
	
	\subfigure[\textit{Item} --- a set of elements over the domain $D = \{ d_1, d_2, d_3, d_4, d_5, d_6, d_7 \} $. The items $a$, $b$, $c$ and $d$ are presented. We can always represent a set over a domain $D$ as a binary string of length $|D|$ where the $i$-th bit equals one iff the element $d_i$ is contained in the set.]{\label{fig:term:item} \includegraphics[width=0.31\textwidth]{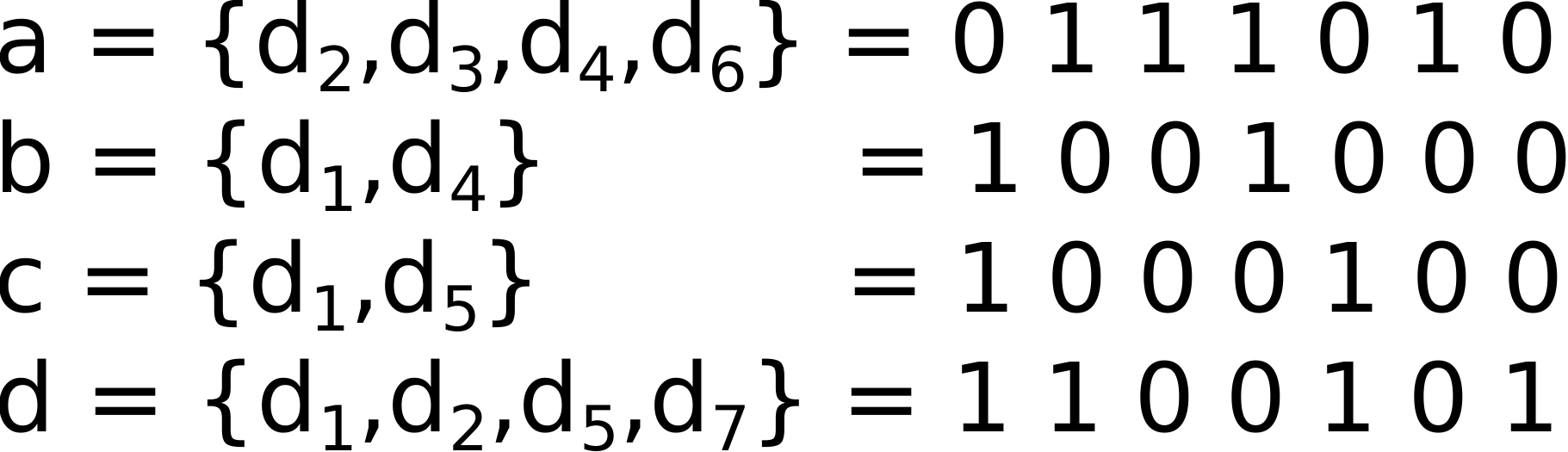}}
	~~~~
	\subfigure[\textit{Itemset} --- set of items over the domain $D$. The two itemsets $S = \{b,d\}$ and $T = \{a,c,d\}$ over the domain $D = \{ d_1, d_2, d_3, d_4, d_5, d_6, d_7 \} $ are presented. Remember that there are no duplicate items within an itemset.]{\label{fig:term:itemset} \includegraphics[width=0.55\textwidth]{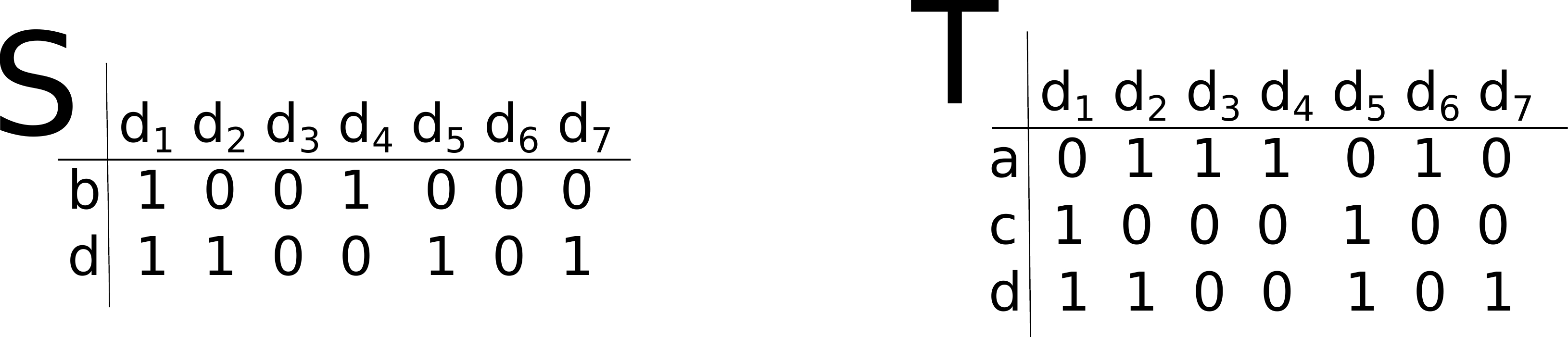}}
	~~~~~~~~~~~~~~~~~~~~~~~~~~~~~~~~~~~~~~~~~~`
	\subfigure[\textit{Dataset} --- ordered set of itemsets over the domain $D$. The dataset $F = \langle S, T \rangle$ over the domain $D = \{ d_1, d_2, d_3, d_4, d_5, d_6, d_7 \} $ is presented. Remember that the itemsets within a dataset are ordered increasingly by cardinality.]{\label{fig:term:dataset} \includegraphics[width=0.67\textwidth]{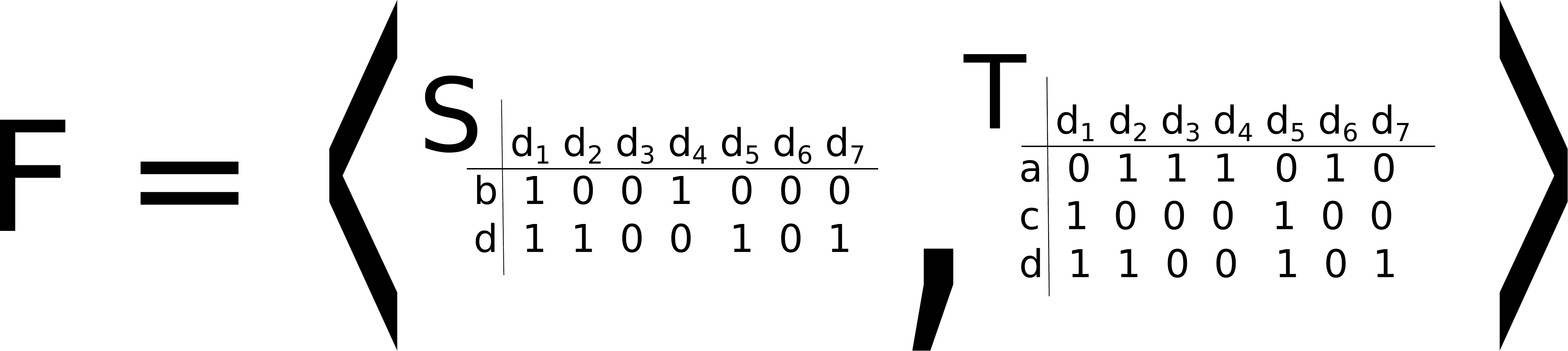}}

	\caption{Graphical representation of the mathematical objects that are used throughout the paper --- item, itemset and dataset. For all of the examples in this figure, we use a dataset $D = \{d_1, d_2, \dots, d_7\}$ of seven elements. For a formal definition, please refer to Section~\ref{sec:intro:term}.}
	
	\label{fig:term}
\end{figure}

We now precede with the formal definitions all of the mathematical objects that are used throughout this paper. Visual examples of all object types are presented in Figure~\ref{fig:term}. Unless otherwise stated, we assume to be working in the domain $D = \{ d_1, d_2, \dots, d_n \}$ of $n$ distinct \textit{elements}.

\begin{itemize}
\item \textit{item} --- a set of elements over the domain $D$. We represent an item $I$ as a binary string of length $n$ where the $i$-th bit is equal to $1$ iff the element $d_i \in I$ for all $1 \leq i \leq n$; i.e. $I \subseteq \{0,1\}^n$. See Figure~\ref{fig:term:item} for examples of items.

\item \textit{itemset} --- a set of items over the domain $D$. We represent an itemset $S$ as a matrix with $|S|$ rows and $n$ columns over the field $\{0,1\}$. See Figure~\ref{fig:term:itemset} for examples of itemsets.

\item \textit{dataset} --- an ordered set of itemsets by cardinality in ascending order over the domain $D$. See Figure~\ref{fig:term:dataset} for examples of datasets.

\end{itemize}

We have chosen the labels of the objects to match that of itemset mining algorithms \cite{Pritchard97} \cite{BayardoPanda11} \cite{Fort+14} \cite{Marinov:ExtremalSets} because the extremal sets identification problem is a sub-problem of the main task. Codish~et~al. \cite{CodishCS14:Two_Layer_Prefix} \cite{CodishCFS14_Comparators} describe a subset of our problem as `words up to permutations' instead of the generalization of `itemsets up to subitemset isomorphism'. We consider the choice of naming objects to be personal preference, because all that is important is the mathematical structure of the object that we work with, not the labels used. Hence, we are as rigorous as possible in this Section~\ref{sec:intro:term}, when it comes to defining the objects.


\subsection{Operations}
\label{sec:intro:ops}

Having defined all of the necessary terminology (Section~\ref{sec:intro:term}) we now formally put forward the respective operations that we investigate in this paper.

\begin{definition}
\label{def:iso:itemset}
Let $S$ and $T$ be itemsets over the domains $D_S$ and $D_T$, respectively. We say that $S$ is isomorphic to $T$ iff there exists a bijection $J : D_S \longrightarrow D_T$ such that $J(S) = T$, also written as $S \cong T$; where $J(S) = \{ \{ J(d) ~|~ d \in I \} ~|~ I \in S \}$. If $D_S = D_T$ then we refer to $J$ as an automorphism.
\end{definition}

\begin{remark}
\label{rem:iso:itemset}
Let $S = \{ S_1, S_2, \dots, S_k \}$ and $T  = \{ T_1, T_2, \dots, T_k \}$ be itemsets over the domains $D_S$ and $D_T$ such that $S$ is isomorphic to $T$ given by $J(S) = T$. Since $S$ and $T$ are sets there exists a bijection $\sigma : \{1, 2, \dots k\} \longrightarrow \{1, 2, \dots k\}$ which maps the items from $S$ to the items in $T$ such that $\forall S_i \in S $ we have $J(S_i) = T_{\sigma(i)}$.
\end{remark}

\begin{definition}
\label{def:iso:subitemset}
Let $S$ and $T$ be itemsets over the same domain $D$. We say that $S$ is subset of $T$ up to isomorphism iff there exists a bijective $J : D \longrightarrow D$ such that $J(S) \subseteq T$, also written as $S \preceq T$.
\end{definition}

\subsection{The Problems of Interest}
\label{sec:intro:prob}

\begin{definition}{Itemset Isomorphism (II) decision problem:}
\label{def:subset_perm}

\underline{Input}:
Two itemsets $S$ and $T$ over the domains $D_S$ and $D_T$, respectively.

\underline{Question}:
Is there a bijection $J : D_S \longrightarrow D_T$ s.t. $J(S) = T$?
\end{definition}

\begin{definition}{Subitemset Isomorphism (SI) decision problem:}
\label{def:SI}

\underline{Input}:
Two itemsets $S$ and $T$ over the domain $D$.

\underline{Question}:
Is there a bijection $J : D \longrightarrow D$ s.t. $J(S) \subseteq T$?
\end{definition}

What is the worst-case complexity class of any algorithm for solving the II and SI problems?

\subsection{Contributions}
\label{sec:intro:cont}

The main contributions our work can be summarized as follows.

\begin{itemize}
\item{\emph{Itemset Isomorphism: GI-Complete}} --- In Section~\ref{sec:complexity} we present a proof that the itemset isomorphism decision problem (equality up to bijection of itemsets) is exactly as difficult as the Graph Isomorphism decision problem. The problem is of great importance in the sorting network optimization domain \cite{BundalaZ13_Optimal_Depth}.

\item{\emph{Subitemset Isomorphism: GI-Hard}} --- As an immediate consequence, the problem of finding a class representative itemsets up to subitemset isomorphism within a dataset is GI-Hard, that is at least as hard as GI. This problem has been encountered before in recent research \cite{BundalaCCSZ14_Optimal_Depth} \cite{BundalaZ13_Optimal_Depth} \cite{BundalaZ14_Optimal_Depth} \cite{CodishCFS14_Comparators} \cite{CodishCS14:Two_Layer_Prefix} in the sorting networks optimization domain, but its worst case computational complexity has never been classified.
\end{itemize}

\section{Motivation: Sorting Network Optimization}
\label{sec:motivation}

\subsection{Preliminaries}
A sorting network is a mathematical object consisting of exactly $n$ wires and comparators designed to sort an input of $n$ numbers. Sorting networks are oblivious to the order of the input data and always perform the same set of pre-determined operations to produce a sorted list of $n$ numbers. The problem of finding optimal sorting networks was first proposed \cite{Bose:1962:SP} by Bose and Nelson more than $50$ years ago. There are two common measures for the optimality of a sorting network --- number of levels (depth) and number of comparators. The problem studied in this paper is central \cite{BundalaCCSZ14_Optimal_Depth} \cite{BundalaZ13_Optimal_Depth} \cite{BundalaZ14_Optimal_Depth} \cite{CodishCS14:Two_Layer_Prefix} to both sorting network optimization problems.

\subsection{The Key Concept}
We now restate \cite{BundalaZ13_Optimal_Depth} a key idea in optimizing sorting networks. First, we note that any comparator network can be represented as the itemset of outputs when the network is applied to all possible inputs. The number of possible comparator networks is exponential in the number of channels. Therefore the enormous search space for optimal sorting networks naturally gives rise to the concept of subitemset isomorphism --- denoted as $\preceq$. More specifically, when searching for optimal sorting networks, we can discard comparator networks whose itemset representation is not minimal up to $\preceq$; i.e. given a dataset $D$, we need to find a class representative itemsets up to $\preceq$ within $D$ that are of minimal cardinality. This idea is at the core of recent computer-assisted proves for the level-optimal $n$-input sorting networks for $11 \leq n \leq 17$ \cite{CodishCS14:Two_Layer_Prefix} \cite{BundalaZ13_Optimal_Depth} \cite{BundalaCCSZ14_Optimal_Depth} \cite{BundalaZ14_Optimal_Depth} \cite{Ehlers:X:15}; an algorithm using the same idea was built \cite{CodishCFS14_Comparators} to provide computer-assisted proof for the comparator-optimal $n$-input sorting networks for $10 \leq n \leq 11$. 


\begin{remark}
In this Section~\ref{sec:motivation}, we omit some important details, due to space and topic limitations, like the exact form/shape of the dataset $D$ where we are allowed to make such search space reduction in the sorting networks optimization domain. For more information on this topic, we refer the reader to the excellent papers \cite{CodishCS14:Two_Layer_Prefix} \cite{BundalaZ13_Optimal_Depth} \cite{BundalaZ14_Optimal_Depth} \cite{BundalaCCSZ14_Optimal_Depth} by Bundala, Zavodny and Codish et al. They were the first to put forward the idea of subitemset isomorphism (in the sorting networks optimization domain), although not in such a general context as presented in this paper.
\end{remark}

\section{Related Work}
\label{sec:related}

\subsection{Complexity Analysis of CM \cite{ChoiM02} Sorting Network Isomorphism}

Choi and Moon \cite{ChoiM02} describe a sorting network isomorphism and present an algorithm aimed at reducing the search space in sorting networks optimization. They show that their (CM) isomorphism is polynomial-time equivalent to the Graph Isomorphism problem of bounded valance. The GI problem of bounded valance can be solved efficiently \cite{LUKS:82:GI:Valance} in polynomial time. 

The work of Choi and Moon is an inspiration to our work because we examine a stronger (BZ \cite{BundalaZ13_Optimal_Depth}) isomorphism of sorting networks than the CM isomorphism. We prove that the BZ itemset isomorphism problem is polynomial-time equivalent to the (generic) version of the Graph Isomorphism problem. Moreover, we show that the BZ subitemset isomorphism problem is GI-Hard.

\subsection{Known Algorithms for the Subitemset Isomorphism Problem}


Given a dataset $F$, the relation $\preceq$ induces a partial order on $F$. To optimize the search for optimal sorting networks it is enough \cite{CodishCFS14_Comparators}(Section 3) for one to consider only the minimal representative itemsets within $F$ up to $\preceq$.

Multiple \cite{Marinov:ExtremalSets:Permutation} \cite{CodishCFS14_Comparators} \cite{BundalaZ13_Optimal_Depth}, very fast deterministic practical algorithms for finding a class representative of minimal up to $\preceq$ itemsets within a dataset $D$ exist in the literature. However, the worst case complexity of all these known algorithms is exponential in $n$ --- the size of the domain/alphabet, as defined in Section~\ref{sec:intro:term}.

In this paper, we narrow the gap between theory and practice by formally proving that the SI decision problem is at least as difficult as the Graph Isomorphism (GI) decision problem. Moreover, we show that the Itemset Isomorphism problem is GI-Complete; that is polynomial-time equivalent to the Graph Isomorphism (GI) problem.

\subsection{Complexity Analysis of Graph Isomorphism}

The graph isomorphism problem is one of two listed~\cite{GareyJ79:Theory:NP-Completeness} by Garey and Johnson that is yet to be classified. The possible complexity classes include but are not limited to: P, NP-Complete, QP (as very recent research suggests by L. Babai). Over the years, there is substantial research on the GI problem: fast practical algorithms with or without domain restrictions \cite{Bodlaender90:GI:Restricted_Domain} \cite{GazitR90:GI:Restricted_Domain} \cite{WagnerDLNT09:GI:Restricted_Domain} , complexity analysis \cite{KoblerST92:LWPP} \cite{Toda99:GI:Complexity} \cite{Johnson85thenp-completeness} \cite{LUKS:82:GI:Valance}, GI-Complete problems \cite{Booth1979problems}  \cite{Zemlyachenko85:GI:Hyper}, etc. More importantly, it is commonly believed that GI-Complete problems form a uniquely defined complexity class that sits between P and NP-Complete, but this is yet to be proved. In this paper, we use the fact that the Hypergraph Isomorphism (HGI) decision problem is polynomial-time equivalent \cite{Zemlyachenko85:GI:Hyper} to the Graph Isomorphism (GI) problem.

\subsection{Itemset Mining}
It is worth noting that the itemset terminology used throughout this paper is widely used in the context of database mining and frequent itemset mining. The graph isomorphism problem correlation to itemset mining is evident in the work of Juan et al. \cite{huan2003efficient} who investigate the subgraph mining problem. Another data mining example is the work of Nuyoshi et al. \cite{miyoshi2009frequent} who use quantitative itemset mining techniques to mine frequent graph patterns. 

Lastly, we note that the problem of finding minimal itemsets within a dataset up to subitemset isomorphism is a generalization of the extremal sets problem \cite{Marinov:ExtremalSets:Permutation}; where in the extremal sets problem relabelling of domain elements is not permitted. The problem of finding extremal sets \cite{BayardoPanda11} \cite{Fort+14} \cite{Marinov:ExtremalSets} has received large attention in recent years. Moreover, practical algorithms \cite{Marinov:ExtremalSets:Permutation} for finding minimal itemsets up to subitemset isomorphism  within a dataset, first find the minimal itemsets within the dataset to reduce the search space.

\section{Complexity Classification}
\label{sec:complexity}


In this section we first formally define the Graph Isomorphism (GI) decision problem \cite{GareyJ79:Theory:NP-Completeness} \cite{Booth1979problems}. We then precede with an informal discussion of how the GI and II problem ``differ''. In Section~\ref{sec:complexity:GI<II} we show that GI $\leq_P$ II and then in Section~\ref{sec:complexity:II<GI} we show that II $\leq_P$ GI. Following is a natural deduction that II is GI-Complete and also the natural consequence of the GI-Hardness of SI (GI $\leq_P$ SI).


\begin{definition}{Graph Isomorphism (GI) decision problem:}
\label{def:graph_iso}

\underline{Input}: 
Two undirected graphs $G = \langle V_G, E_G\rangle$ and $H = \langle V_H, E_H\rangle$. 

\underline{Question}:
Is there a bijection $I : V_G \longrightarrow V_H$ s.t. $(v,w) \in E_G$ iff $(I(v), I(w)) \in E_H$?
\end{definition}

\subsection{Discussion}
\label{sec:complexity:discuss}

\begin{figure} [t]
	\centering
	\includegraphics[width=0.731\textwidth]{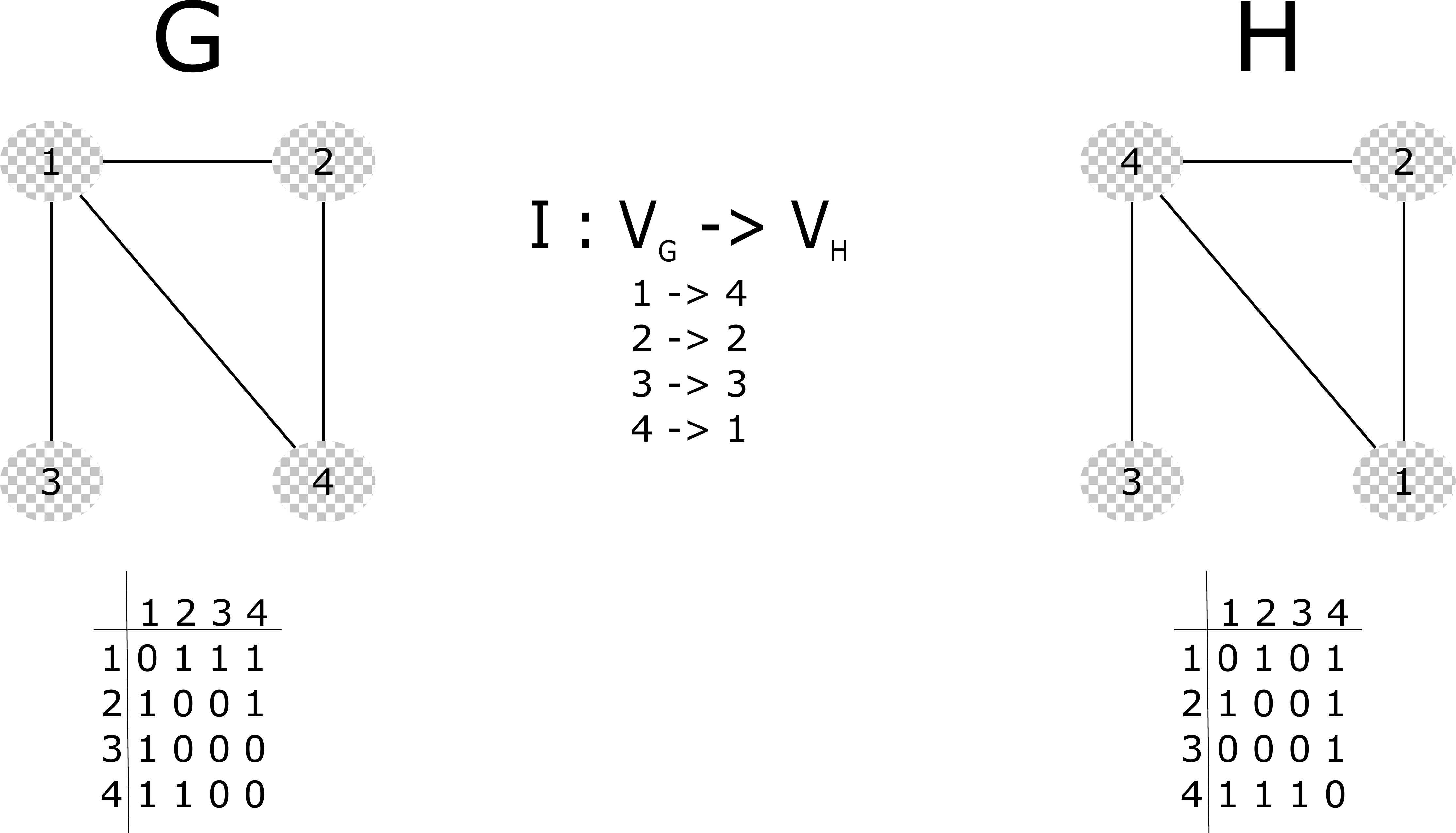}
	\caption{This figure presents a graph $G$ and its adjacency matrix. We also present a swap of the vertices $1$ and $4$ of $G$ to obtain $H$. To construct the adjacency matrix of $H$ from the adjacency matrix of $G$, we first swap the rows $1$ and $4$ of $G$ and then swap the columns $1$ and $4$. Since, every permutation (of the vertices of $G$) can be written as a sequence of swaps, this figure shows the methodology of applying a permutation to a graph. }
	\label{fig:GI}
\end{figure}

\begin{figure} [t]
	\centering
	\includegraphics[width=0.731\textwidth]{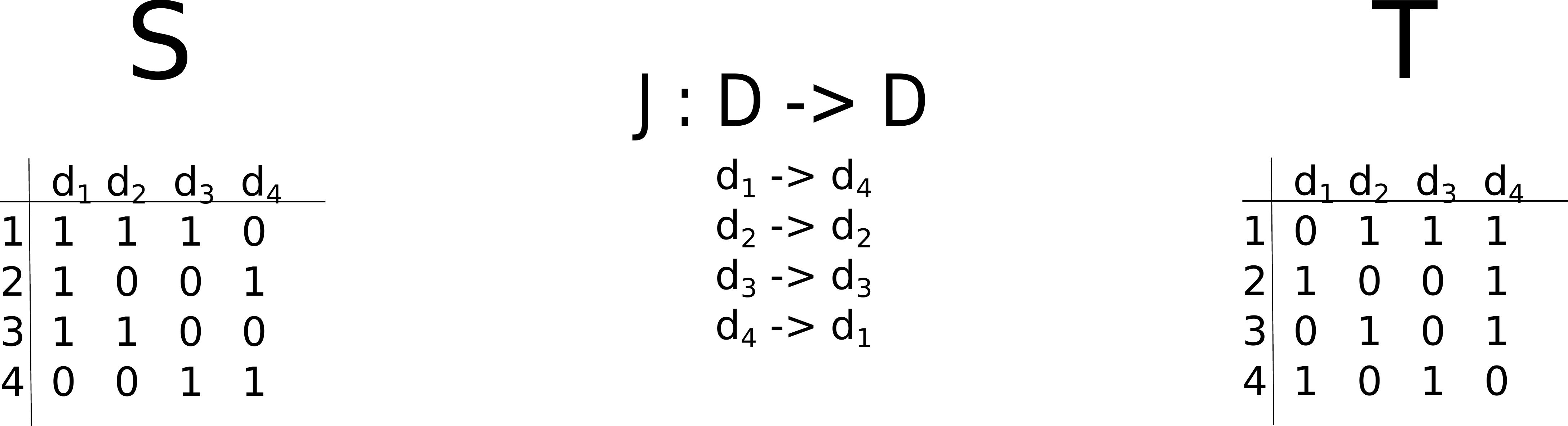}
	\caption{This figure presents an itemset $S$ over the domain $D = \{d_1, d_2, d_3, d_4\}$ and its matrix representation (as described in Section~\ref{sec:intro:term} and Figure~\ref{fig:term:itemset}). We also present a swap of the domain elements $d_1$ and $d_4$ of $S$ to obtain $T$. To construct the matrix representation of $T$ from the  matrix of $S$, we need to swap the two columns $d_1$ and $d_4$. Since, every permutation (of the domain $D$) can be written as a sequence of swaps, this figure shows the methodology of applying a permutation to an itemset. }
	\label{fig:II}
\end{figure}

Before presenting a rather technical proof that the II decision problem is GI-Complete, we give a brief discussion on how the GI and SI problems ``differ''. Intuitively, the two problems are very similar as the inputs to both problems can be represented as zero-one matrices --- see Figures~\ref{fig:GI} and \ref{fig:II}; however, there are two fundamental differences.

\begin{itemize}
\item In the GI problem a swap of vertices is represented as a swap of two rows and two columns of the zero-one adjacency matrix (Figure~\ref{fig:GI}), whereas in the II problem a swap of two domain elements is represented as a swap of two columns (Figure~\ref{fig:II}) with the rows left intact.

\item A valid solution for the GI problem requires the two zero-one adjacency matrices to match exactly. Whereas, in the II problem any reordering of the rows is permitted (recall Remark~\ref{rem:iso:itemset}).
\end{itemize}

\subsection{II is GI-Hard}
\label{sec:complexity:GI<II}

\begin{lemma}
\label{lem:GI<II}
GI $\leq_P$ II.
\end{lemma}

The proof of Lemma~\ref{lem:GI<II} is a rather technical one. However, the proof is constructive, and we present examples in Figures \ref{fig:examples:GI<II:A} and \ref{fig:examples:GI<II:B} for the essential steps of the proof. A detailed explanation of the examples following the steps of the proof of Lemma~\ref{lem:GI<II} is presented in \ref{sec:examples:GI<II}.

\begin{proof}
\label{proof:lem:graph_iso<itemset_iso}
Define the function $f : \langle G, H \rangle = \langle S, T \rangle$ where $\langle G, H \rangle$ is input to GI and $\langle S, T \rangle$ is an input to II. The itemset $S = \{ S_u ~|~ u \in V_G \}$ where the items $S_u = \{ (u,v) \in E_G ~|~ v \in V_G \}$. Similarly, the itemset $T = \{ T_h ~|~ h \in V_H \}$ where the items $T_h = \{ (h,w) \in E_H ~|~ w \in V_H \}$. We now show that the function $f$ is a polynomial-time reduction of Graph-Isomorphism to Itemset-Isomorphism.

First, we need to show that the function $f$ is a polynomial time one. It is obvious, that this is the case, because $f$ does no computation and simply, re-structures the input. Hence, the reduction function $f$ is polynomial time.

To prove that the presented polynomial-time reduction is correct, we need to show that a Graph-Isomorphism instance is satisfiable (yes instance), if and only if the $f$-induced Itemset-Isomorphism  instance is  satisfiable.

Suppose that the Graph-Isomorphism instance is satisfiable: there exists a bijection $I : V_G \longrightarrow V_H$ s.t. $(v,w) \in E_G$ iff $(I(v), I(w)) \in E_H$. We claim that $J : (v,w) \longrightarrow (I(v), I(w))$ satisfies $J(S) = T$. To see this, consider any item $S_g = \{ (g,x) \in E_G ~|~ x \in V_G \} \in S$ and apply the bijection $J$ to it. Then clearly we have $J(S_g) = \{ (I(g), I(x)) \in V_H ~|~ x \in V_G \} =  \{ (I(g), y) \in V_H ~|~ y \in V_H \} = T_{I(g)} \in T$. Also note that, since $I$ is bijective then $I^{-1}$ exists and we can similarly show that for any $T_h \in T$ we have $I^{-1}(T_h) = S_{I^{-1}(h)} \in S$ . Hence, we have shown that, if any Graph-Isomorphism instance $\langle G,H \rangle$ is satisfiable then the created Itemset-Isomorphism instance $f(\langle G,H \rangle)$ is satisfiable.

Now suppose that the created Itemset-Isomorphism instance $f(\langle G,H \rangle) = \langle S,T \rangle$ is satisfiable: there is a bijection $J : E_G \longrightarrow E_H$ s.t. $J(S) = T$. By Definition~\ref{def:iso:itemset} of itemset isomorphism and Remark~\ref{rem:iso:itemset}, we know there exists a bijection $\sigma$ that maps the items in $J(S)$ to the items in $T$. Hence, $\sigma : V \longrightarrow H$ is such that for any $S_g \in S$ we have $ J(S_g) = T_{\sigma(g)} \in T$, and vice versa. We claim that $\sigma$ gives a graph isomorphism from $G$ to $H$. To see this, notice that for all $(v,w) \in E_G$ we have $(\sigma(v), \sigma(w)) = J((v,w))$ (we are working with undirected graphs). But, from the assumption we know that $J((v,w)) \in E_H$; to go from $E_H$ to $E_G$ is systematically the same because $\sigma$ is bijective, hence $\sigma^{-1}$ exists. Therefore, we have shown that if the created Itemset-Isomorphism instance $f(\langle G,H \rangle) = \langle S,T \rangle$ is satisfiable then the original Graph-Isomorphism instance $\langle G,H \rangle$ is satisfiable.
\end{proof}

\subsection{GI is II-Hard}
\label{sec:complexity:II<GI}

In order to prove that there is polynomial-time reduction from the II decision problem to the GI decision problem, we require an intermediate result. Namely, we are going to show that the Hypergraph Isomorphism (HGI, see the following Definition~\ref{def:HGI}) decision problem is II-Hard; in other words, we will show that HGI is at least as hard (up to a polynomial transformation) as II. We also use the known fact that HGI is polynomial-time equivalent to GI, formally stated in Theorem~\ref{th:GI=HGI}. 

We refer to a hypergraph $G = \langle V_G, E_G\rangle$ as a combinatorial object consisting of nodes $V_G$ and edges $E_G$ except that the edges consist of an arbitrary number of nodes within $V_G$ \cite{Lawler73:Hypergraphs}. Examples of hypergraphs are given in \ref{sec:examples:II<HGI}.

\begin{definition}{Hypergraph Isomorphism (HGI) decision problem:}
\label{def:HGI}

\underline{Input}: 
Two undirected hypergraphs $G = \langle V_G, E_G\rangle$ and $H = \langle V_H, E_H\rangle$.

\underline{Question}:
Is there a bijection $I : V_G \longrightarrow V_H$ s.t. $A \in E_G$ iff $I(A) \in E_H$?

where $A \subseteq V_G$ and $I(A) = \{ I(v) ~|~ v \in A \}$.
\end{definition}

\begin{theorem}
\label{th:GI=HGI}
HGI is GI-Complete.
\end{theorem}

\begin{proof}
The statement of Theorem~\ref{th:GI=HGI} is very well known in the community \cite{babai2008isomorhism}. We refer the reader to \cite{Zemlyachenko85:GI:Hyper}.
\end{proof}

To summarize our overall strategy, we prove that II $\leq_P$ HGI but since HGI is GI-Complete, we deduce that GI is II-Hard.

\begin{lemma}
\label{lem:II<HGI}
II $\leq_P$ HGI.
\end{lemma}

The proof of Lemma~\ref{lem:II<HGI} is very similar to the proof of Lemma~\ref{lem:GI<II} from Section~\ref{sec:complexity:GI<II}. However, there is a complication: namely, given an instance $\langle S,T \rangle$ of the II problem such that either $S$ or $T$ contain a column with total number of ones not equal to two, in the matrix representation (as described in Section~\ref{sec:intro:term} and Figure~\ref{fig:term:itemset}). This situation requires a slightly different mechanism to the one described in the proof of Lemma~\ref{lem:GI<II}.

In the following proof of Lemma~\ref{lem:II<HGI} we define a polynomial-time reduction function $g : \langle S,T \rangle \longrightarrow \langle G,H \rangle$ from the II problem to the HGI problem. The natural extension of our machinery (proof of Lemma~\ref{lem:GI<II}) to solve the complication is to think of the ones in any column of the matrix representation of an itemset as defining a hyperedge in a hypergraph. As an example, consider the itemset $S$ in Figure~\ref{fig:II}; then the $g$-corresponding hyperedge for the column $d_1$ would be the set of nodes $\{1, 2, 3\}$, and the hyperedge for the column $d_2$ would be the set of nodes $\{1,3\}$ --- this is a normal edge only because the cardinality of the hyperedge is 2. We give examples of the following constructive proof in \ref{sec:examples:II<HGI}. 

\begin{proof}
\label{proof:lem:II<HGI}
Let us start by formally defining the function $g : \langle S,T \rangle \longrightarrow \langle G,H \rangle$ over the itemsets $S = \{ S_g \}$ and $T = \{ T_h \}$ over the domains $D_S = \{s_1, s_2, \dots, s_n\}$ and $D_T = \{t_1, t_2, \dots, t_n\}$ respectively. We set $G = \langle V_G,E_G \rangle$ and $H = \langle V_H,E_H \rangle$ s.t. $V_G = \{ g ~|~ S_g \in S \}$, $E_s = \{ g ~|~ s \in S_g : S_g \in S\}$, $E_G = \{ E_s ~|~ s \in D_S \}$ and $V_H = \{ h ~|~ T_h \in T \}$, $E_t = \{ h ~|~ t \in T_h : T_h \in T\}$, $E_H = \{ E_t ~|~ t \in D_T \}$. We claim that $g$ is a polynomial-time reduction of II to HGI.

It is clear that $g$ can be implemented in a polynomial time, since $g$ performs no calculation and transforms the itemsets $S$ and $T$ to the hypergraphs $G$ and $H$ respectively; where $|V_G| = |S|$, $|E_G| = |D_S|$, $|V_H| = |T|$ and $|E_H| = |D_T|$. 

It is now left to show that the Itemset Isomorphism instance $\langle S,T \rangle$ is satisfiable (yes instance) if and only if the $g$-induced Hypergraph Isomorphism instance $\langle G,H \rangle$ is satisfiable.

Suppose that the Itemset Isomorphism instance $\langle S,T \rangle$ is satisfiable: there exists a bijection $J : D_S \longrightarrow D_T$ s.t. $J(S)=T$. Using Definition~\ref{def:iso:itemset}, Remark~\ref{rem:iso:itemset} and the construction of $g$, we deduce there exists a bijection $\sigma : V_G \longrightarrow V_H$ s.t. $S_g \in S$ if and only if $J(S_g) = T_{\sigma(g)} \in T$. We claim that the bijection $\sigma$ gives a hypergraph isomorphism between $G$ and $H$. To prove our claim, consider any $E_s = \{ g ~|~ s \in S_g : S_g \in S\} \in E_G$. However, applying the bijections $\sigma$ and $J$ to the set of vertices $E_s$ (hyperedge) is equivalent to $ \{ \sigma(g) ~|~ J(s) \in J(S_g) : J(S_g) \in J(S) \} $; simplifying (by setting $h = \sigma(g)$ and $J(S_g) = T_h$) gives us $\{ h ~|~ t \in T_h : T_h \in T \} = E_{J(s)} = E_{h}$. Since, $\sigma$ and $J$ are bijective, we deduce that $\sigma$ gives a hypergraph isomorphism between $G$ and $H$. Hence, we have shown that if the Itemset Isomorphism instance $\langle S,T \rangle$ is satisfiable then the Hypergraph Isomorphism instance $ g(\langle S,T \rangle)=\langle G,H \rangle$ is also satisfiable.

Suppose that the Hypergraph Isomorphism instance $ g(\langle S,T \rangle)=\langle G,H \rangle$ is satisfiable: there exists a bijection $I : V_G \longrightarrow V_H$ s.t. $E_s \in E_G$ if and only if $I(E_s) \in E_H$. However, since $E_H = \{ E_t ~|~ t \in D_T \}$ then we can define a bijection $\gamma : D_S \longrightarrow D_T$ such that $s \longmapsto t$ if and only if $I(E_s) = E_t$. We claim that $\gamma$ gives an Itemset Isomorphism between $S$ and $T$. This is trivially seen because $\gamma( S_g ) = T_{I(g)}$  for all $g \in V_G$ and due to $\gamma$ and $I$ being bijections.
\end{proof}

\begin{lemma}
\label{lem:II<GI}
II $\leq_P$ GI.
\end{lemma}

\begin{proof}
\label{proof:lem:II<GI}
Immediate consequence of applying Theorem~\ref{th:GI=HGI} and Lemma~\ref{lem:II<HGI}.
\end{proof}

\subsection{II is GI-Complete}
\label{sec:complexity:II=GI}

To complete our proof that II is GI-Complete, we still need to show that II $\in$ NP.

\begin{lemma}
	\label{lem:itemset_iso:NP}
	II $\in$ NP.
\end{lemma}

\begin{proof}
	\label{proof:lem:itemset_iso:NP}
	We need to show that a polynomial time verifier of the Itemset-Isomorphism problem exists to conclude that II is in NP. It is easy to see that, given a bijection $J$ the verifier needs to check if $J(S) = T$. Clearly the application of the bijection $J$ to $S$ can be done in polynomial time. The equality checking can be done in polynomial time because $J(S)$ and $T$ are sets of a polynomial number of elements each.
\end{proof}

\begin{theorem}
\label{th:gi-hard:itemset_iso}
II is GI-Complete.
\end{theorem}

\begin{proof}
\label{proof:th:gi-complete:itemset_iso}
Follows immediately by applying Lemmas~\ref{lem:GI<II}, \ref{lem:II<GI} and \ref{lem:itemset_iso:NP}.
\end{proof}

\begin{corollary}
\label{corr:gi-complete:itemset_iso:subitemset_iso}
SI is GI-Hard.
\end{corollary}

\begin{proof}
An immediate consequence to Theorem~\ref{th:gi-hard:itemset_iso} because obviously II $\leq_P$ SI and SI $\in$ NP.
\end{proof}

Furthermore, we deduce that finding a class representative \cite{BundalaZ13_Optimal_Depth} \cite{BundalaZ14_Optimal_Depth} \cite{CodishCFS14_Comparators} up to subitemset isomorphism is GI-Hard --- clearly polynomial-time reducible to the SI problem.

\section{Conclusion and Future Work}
Fast algorithms for the Subitemset Isomorphism (SI) problem are of practical importance in the sorting networks optimization domain. The SI problem is encountered in recent \cite{BundalaCCSZ14_Optimal_Depth} \cite{BundalaZ14_Optimal_Depth} \cite{BundalaZ13_Optimal_Depth} \cite{CodishCFS14_Comparators} breakthrough sorting networks optimization research however its worst-case computational complexity classification is an open problem. This current paper proves the Itemset Isomorphism (II) decision problem to be GI-Complete; polynomial-time equivalent to the Graph Isomorphism (GI) decision problem. As a corollary, the SI problem is shown to be GI-Hard. The complexity analysis presented here is of importance to research aimed at fast practical algorithms \cite{CodishCFS14_Comparators} \cite{Marinov:ExtremalSets:Permutation} for the SI problem, as well as, extending the list \cite{Booth1979problems} of GI-Complete problems which are of practical importance too.

For future work, we aim to classify the SI problem more precisely, rather than the lower complexity bound given here. We suspect, that the problem of Subitemset Isomorphism is NP-Complete. The reason, there is an intuitive relation between the pair of Graph-Isomorphism (GI-Complete) and the Subgraph-Isomorphism (NP-Complete) and the pair of Itemset-Isomorphism (GI-Complete) and Subitemset-Isomorphism (GI-Hard).

\section{Acknowledgements}

Work supported by the Irish Research Council (IRC) and Science Foundation Ireland grant 12/IA/1381.

\bibliographystyle{elsarticle-num}
\bibliography{Manuscript}

\appendix

\clearpage
\section{Examples: II is GI-Hard}
\label{sec:examples:GI<II}
The examples presented in Figures \ref{fig:examples:GI<II:A} and \ref{fig:examples:GI<II:B} demonstrate how to apply the polynomial-time transformation function $f$ (defined in the proof of Lemma~\ref{lem:GI<II}) to an instance $\langle G,H \rangle$ of the GI problem to produce an instance $\langle S,T \rangle$ of the II problem. From the examples, it is clear that $\langle G,H \rangle$ is satisfiable if and only if $\langle S,T \rangle$ is satisfiable.

Following the proof of Lemma~\ref{lem:GI<II} and the two Figures \ref{fig:examples:GI<II:A} and \ref{fig:examples:GI<II:B}, we see exactly how to construct $J$ using $I$, and vice versa; where $I$ gives the graph isomorphism between $G$ and $H$, and $J$ gives the itemset isomorphism between $S$ and $T$. Note, that Lemma~\ref{lem:GI<II} works only for undirected graphs; a more technical proof is required for the case of directed graphs but is not necessary for the complexity classification of the itemset isomorphism problem.

\begin{figure} [t]
	\centering
	\includegraphics[width=0.731\textwidth]{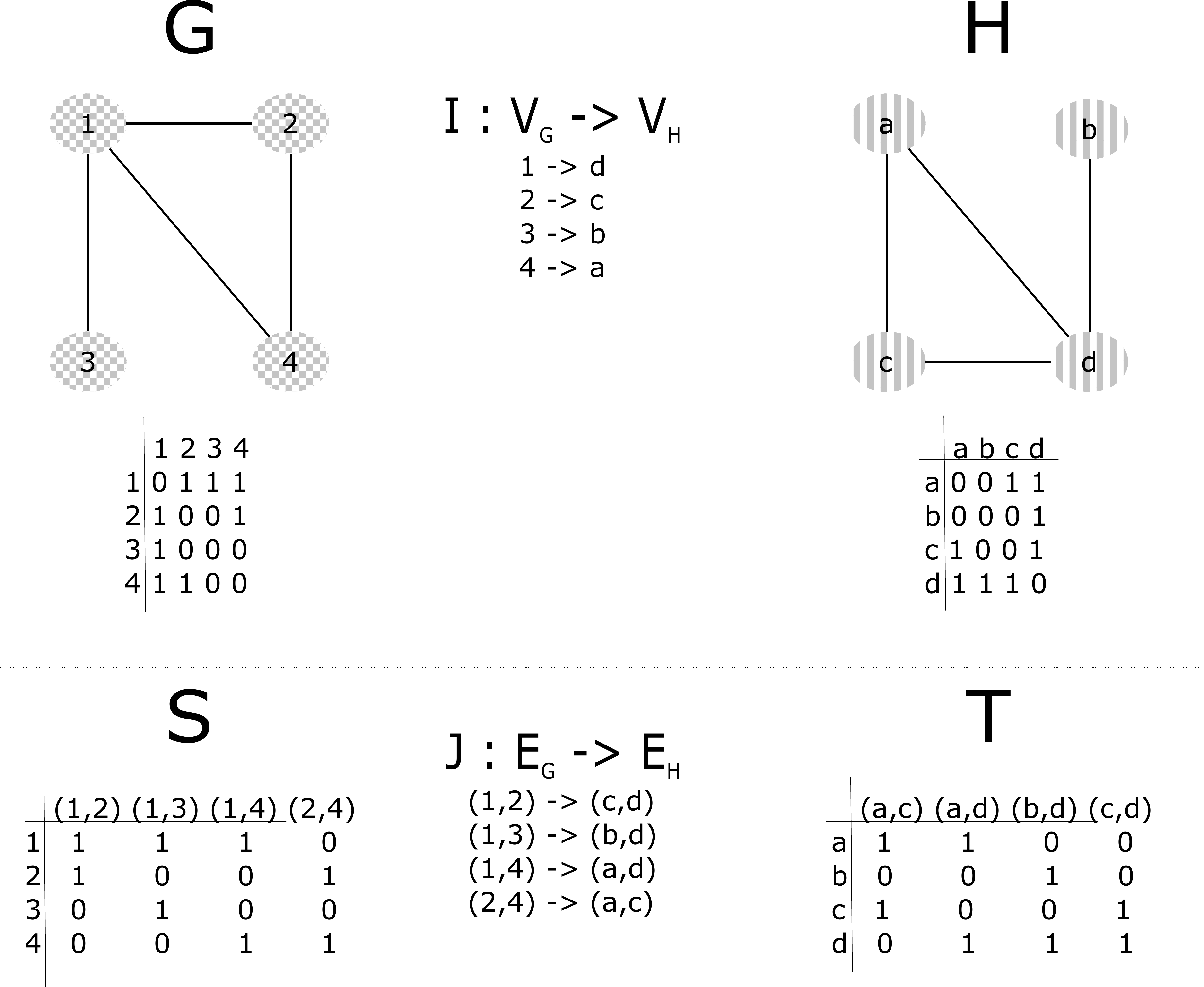}
	\caption{An example of two isomorphic graphs $G$ and $H$ together with the corresponding isomorphic itemsets $S$ and $T$ generated by the polynomial-time reduction function $f : \langle G, H \rangle = \langle S, T \rangle$, as described in the proof of Lemma~\ref{lem:GI<II}. This figure serves as a detailed example of the constructive proof to Lemma~\ref{lem:GI<II}. In the figure we see that, there is a unique isomorphism between $G$ and $H$, given by $I$; and a unique isomorphism $J$ between $S$ and $T$. For detailed explanation of this figure refer to Section~\ref{sec:examples:GI<II:A}.}
	\label{fig:examples:GI<II:A}
\end{figure}

\subsection{Figure~\ref{fig:examples:GI<II:A}}
\label{sec:examples:GI<II:A}

The example presented in Figure~\ref{fig:examples:GI<II:A} shows two isomorphic graphs and their $f$-corresponding isomorphic itemsets; recall $f$ from proof of Lemma~\ref{lem:GI<II}. It is clear that, the two graphs $G$ and $H$ are uniquely isomorphic --- there exists a unique bijection $I : V_G \longrightarrow V_H$ that satisfies $(v,w) \in E_G \longleftrightarrow (I(v), I(w)) \in E_H$.

Hence, given the satisfiable instance $\langle G, H \rangle$ and the bijection $I : V_G \longrightarrow V_H$, in the proof of Lemma~\ref{lem:GI<II} we claim that $J : (v,w) \longrightarrow (I(v), I(w))$ satisfies $J(S) = T$. One can easily check the graphs and itemsets in Figure~\ref{fig:examples:GI<II:A} to verify the correctness of this claim.

Now, suppose we are given a bijection $J : E_G \longrightarrow E_H$ s.t. $J(S) = T$. Clearly for the example in Figure~\ref{fig:examples:GI<II:A}, we have a unique $\sigma = I$ that maps the items in $J(S)$ to the items in $T$.

\begin{figure} [t]
	\centering
	\includegraphics[width=0.731\textwidth]{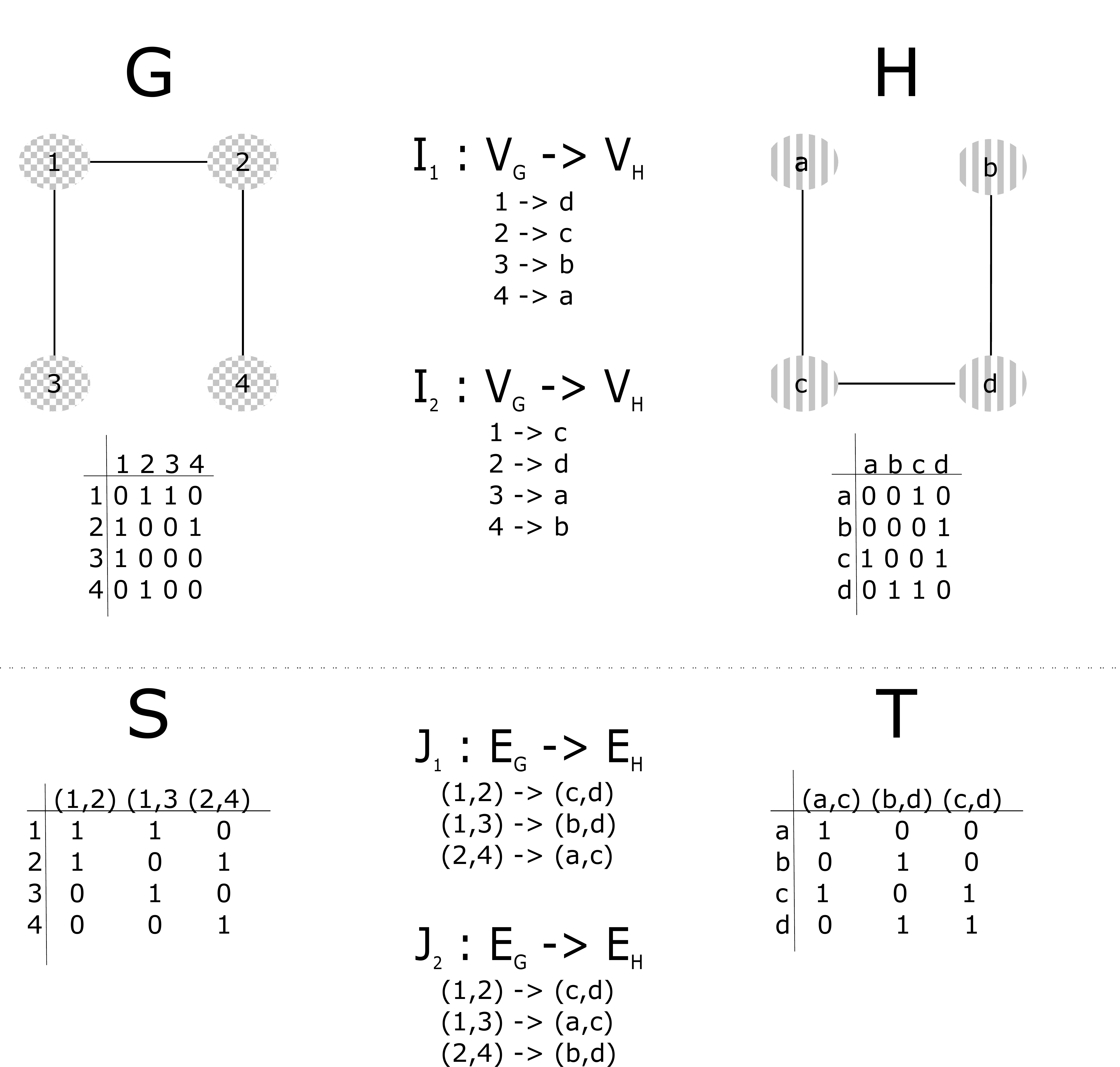}
	\caption{An example of two isomorphic graphs $G$ and $H$ together with the corresponding isomorphic itemsets $S$ and $T$ generated by the polynomial-time reduction function $f : \langle G, H \rangle = \langle S, T \rangle$, as described in the proof of Lemma~\ref{lem:GI<II}. This figure serves as a detailed example of the constructive proof to Lemma~\ref{lem:GI<II}. In the figure we see that, there are exactly two isomorphisms between $G$ and $H$, given by $I_1$ and $I_2$; and exactly two isomorphisms $J_1$ and $J_2$ between $S$ and $T$, where $I_1$ corresponds to $J_1$ and $I_2$ corresponds to $J_2$. For an in-depth explanation of this figure refer to Section~\ref{sec:examples:GI<II:B}.}
	\label{fig:examples:GI<II:B}
\end{figure}

\subsection{Figure~\ref{fig:examples:GI<II:B}}
\label{sec:examples:GI<II:B}

The example presented in Figure~\ref{fig:examples:GI<II:B} shows two isomorphic graphs and their $f$-corresponding isomorphic itemsets. This example is more explanatory than the one presented in Figure~\ref{fig:examples:GI<II:B} because the isomorphisms between the graphs $G$ and $H$ are not unique. Using the constructive proof of Lemma~\ref{lem:GI<II}, we see that the graph isomorphism $I_1$ corresponds to the itemset isomorphism $J_1$ and similarly, $I_2$ corresponds to $J_2$.

\clearpage
\section{Examples: HGI is II-Hard}
\label{sec:examples:II<HGI}
The examples presented in Figures \ref{fig:examples:II<HGI:A} and \ref{fig:examples:II<HGI:B} demonstrate how to apply the polynomial-time transformation function $g$ (defined in the proof of Lemma~\ref{lem:II<GI}) to an instance $\langle S,T \rangle$ of the SI decision problem to produce an instance $\langle G,H \rangle$ of the GI decision problem. From the examples, it is clear that $\langle S,T \rangle$ is satisfiable if and only if $\langle G,H \rangle$ is satisfiable. Note that we focus our attention on itemsets $S$ and $T$ having at least one column with total number of $1$'s not equal to two (a proper hyperedge) in their matrix representation (see Section~\ref{sec:intro:term}).

\begin{figure} [t]
	\centering
	\includegraphics[width=1\textwidth]{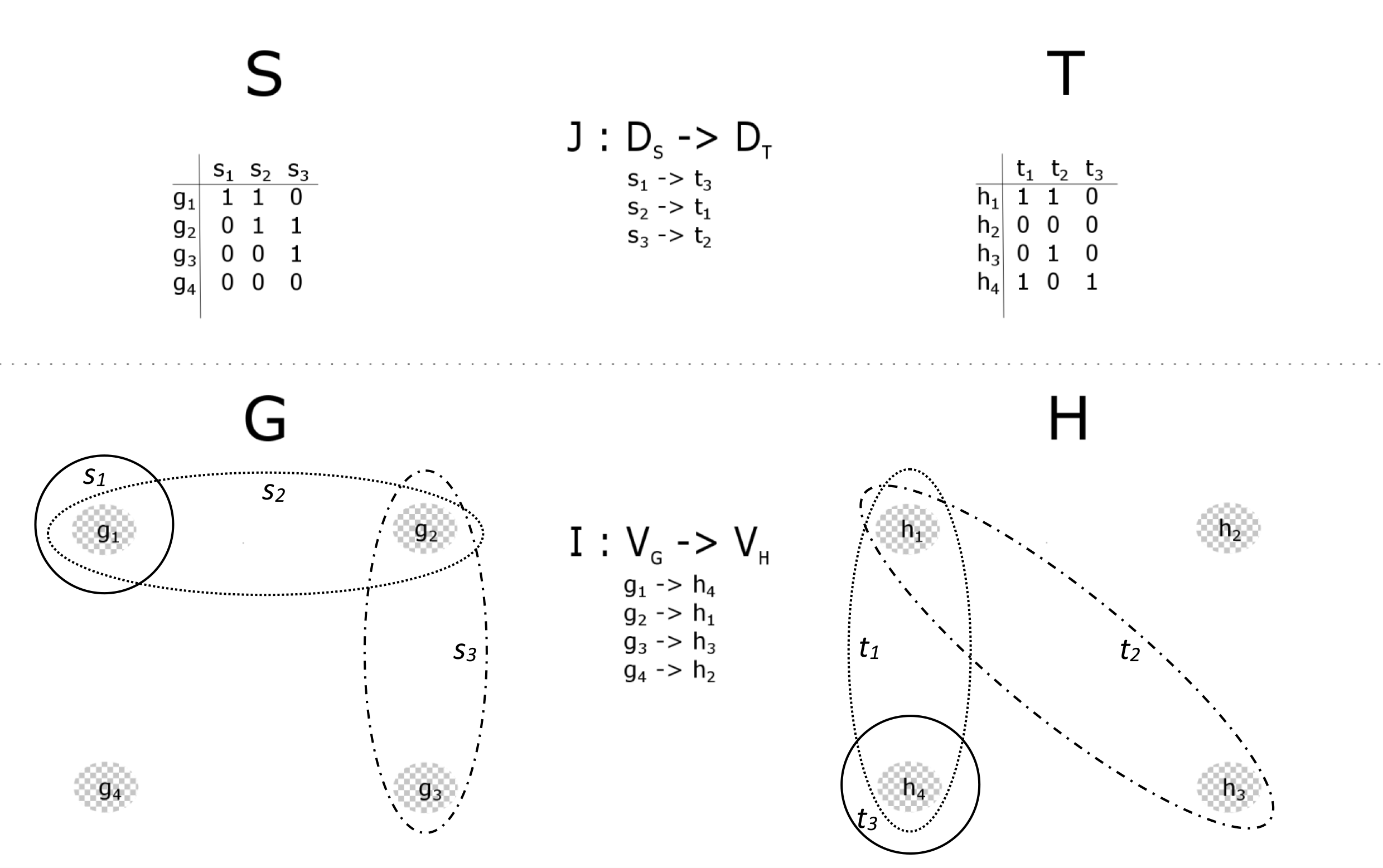}
	\caption{An example of two isomorphic itemsets $S$ and $T$ together with their $g$-corresponding isomorphic hypergraphs $S$ and $T$ generated by the polynomial-time reduction function $g : \langle S, T \rangle = \langle G, H \rangle$, as described in the proof of Lemma~\ref{lem:II<HGI}. This figure serves as a detailed example of the constructive proof to Lemma~\ref{lem:II<HGI}. In the figure we see that, there is a unique isomorphism between $S$ and $T$, given by $J$; and a unique isomorphism $I$ between $G$ and $H$. For detailed explanation of this figure refer to Section~\ref{sec:examples:II<HGI:A}.}
	\label{fig:examples:II<HGI:A}
\end{figure}

\subsection{Figure~\ref{fig:examples:II<HGI:A}}
\label{sec:examples:II<HGI:A}

The purpose of this figure is to gently introduce the reader into itemset and hypergraph representation, as well as to cover the basic idea of the proof of Lemma~\ref{lem:II<HGI}. Figure~\ref{fig:examples:II<HGI:A} shows two uniquely (given by $J$) isomorphic itemsets $S$ and $T$. Following the proof of Lemma~\ref{lem:II<HGI}, we clam that $\sigma = I : V_G \longrightarrow V_H$ gives a graph isomorphism between the $g$-induced hypergraphs $G$ and $H$ from $S$ and $T$ respectively. From the figure, it is clear that $I$ is a unique isomorphism between $G$ and $H$; supporting our claim from the proof of Lemma~\ref{lem:II<HGI} that $G$ and $H$ are isomorphic if and only if $S$ and $T$ are isomorphic, where $g(\langle S,T \rangle) = \langle G,H \rangle$.

\begin{figure} [t]
	\centering
	\includegraphics[width=1\textwidth]{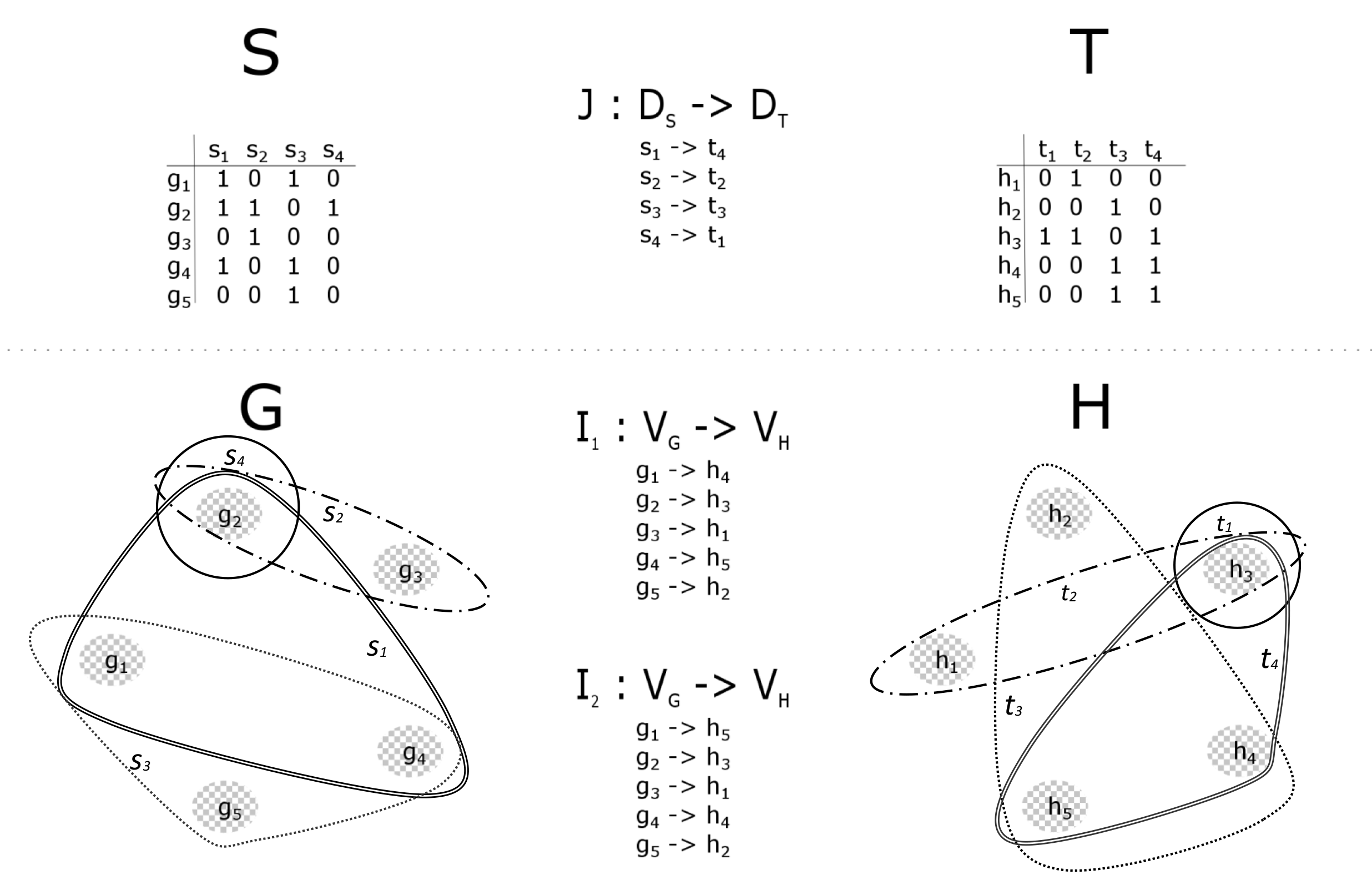}
	\caption{An example of two isomorphic itemsets $S$ and $T$ together with their $g$-corresponding isomorphic hypergraphs $S$ and $T$ generated by the polynomial-time reduction function $g : \langle S, T \rangle = \langle G, H \rangle$, as described in the proof of Lemma~\ref{lem:II<HGI}. This figure serves as a detailed example of the constructive proof to Lemma~\ref{lem:II<HGI}. In the figure we see that, there is a unique isomorphism between $S$ and $T$, given by $J$; and two isomorphisms $I_1$ and $I_2$ between the hypergraphs $G$ and $H$. For detailed explanation of this figure refer to Section~\ref{sec:examples:II<HGI:B}.}
	\label{fig:examples:II<HGI:B}
\end{figure}

\subsection{Figure~\ref{fig:examples:II<HGI:B}}
\label{sec:examples:II<HGI:B}

Figure~\ref{fig:examples:II<HGI:B} shows a rather complicated scenario of two input itemsets $S$ and $T$ and their $g$-induced (as defined in the proof of Lemma~\ref{lem:II<HGI}) graphs $G$ and $H$ respectively. This case is interesting because there is a unique itemset isomorphism $J$ between $S$ and $T$, however there are two ($\sigma_1 = I_1$ and $\sigma_2 = I_2$) graph isomorphisms between $G$ and $H$. Note that the line styles of the hyperedges of both ($G$ and $H$) graphs are matching if and only if one hyperedge is mapped to the other as given by $J$; for example, the hyperedge $s_1$ in $G$ is mapped to the hyperedge $t_4$ in $H$, noting that $J(s_1) = t_4$.

\end{document}